\documentclass[twocolumn,superscriptaddress,showpacs,preprintnumbers,amsmath,amssymb]{revtex4}

\usepackage{epsfig}
\usepackage{dcolumn}% Align table columns on decimal point
\usepackage{bm}% bold math
\usepackage{latexsym}

\begin{document}
%\preprint{APS/123-QED}

\title{Long-lived spin memory in Mn-doped GaAs: Time resolved study}

\author{I.~A. Akimov}
 \affiliation{Experimentelle Physik II, Technische Universit\"at Dortmund, 44221 Dortmund, Germany}
 \affiliation{A.F. Ioffe Physical-Technical Institute, Russian Academy of Sciences, 194021 St. Petersburg, Russia}
\author{R.~I. Dzhioev}
\author{V.~L. Korenev}
\author{Yu.~G. Kusrayev}
 \affiliation{A.F. Ioffe Physical-Technical Institute, Russian Academy of Sciences, 194021 St. Petersburg, Russia}
\author{E.~A. Zhukov}
 \affiliation{Faculty of Physics, M.V. Lomonosov Moscow State University, 119992 Moscow, Russia}
\author{D.~R. Yakovlev}
 \affiliation{Experimentelle Physik II, Technische Universit\"at Dortmund, 44221 Dortmund, Germany}
 \affiliation{A.F. Ioffe Physical-Technical Institute, Russian Academy of Sciences, 194021 St. Petersburg, Russia}
\author{M. Bayer}
 \affiliation{Experimentelle Physik II, Technische Universit\"at Dortmund, 44221 Dortmund, Germany}
\date{\today}

\begin{abstract} We study the electron spin
dynamics in p-type GaAs doped with magnetic Mn acceptors by means of time-resolved pump-probe and photoluminescence techniques. Measurements in transverse magnetic
fields show a long spin relaxation time of 20~ns that can be uniquely related to electrons. Application of weak longitudinal magnetic fields above 100~mT extends
the spin relaxation times up to microseconds which is explained by suppression of the Bir-Aronov-Pikus spin relaxation for the electron on the Mn acceptor.
\end{abstract}

\pacs{78.47.-p/75.50.pp/71.70.Gm/71.35.Ji}

% PACS
%78.47.-p    - Spectroscopy of solid state dynamics
%            78.47.Cd    Time resolved luminescence;  78.47.jm    Quantum beats;  78.47.jc    Time resolved spectroscopy (> 1 psec)
%78.55.Cr    - Photoluminescence, properties and materials / III–V semiconductors
%78.20.Ls    - Optical properties of bulk materials and thin films / Magnetooptical effects
%71.35.Ji    - Excitons and related phenomena /Excitons in magnetic fields; magnetoexcitons
%71.55.Eq    - Impurity and defect levels /    III–V semiconductors
%71.70.Gm    - Level splitting and interactions / Exchange interactions
%75.50.pp    - Studies of specific magnetic materials / Magnetic semiconductors

\keywords{Spin dynamics, Electron spin relaxation, GaAs}

\maketitle

The spin dynamics in semiconductors have attracted strong interest for  several decades and nowadays are in the focus of attention for spin electronics and quantum
information applications \cite{OptOrient, Awschbook, SpinBooks, SpinReview}. Materials with long spin lifetimes and efficient mechanisms for controlling spin
relaxation, e.g. by weak magnetic fields, are required for these purposes. A large amount of work has been mainly done on n-type GaAs, where spin relaxation times
in the range of 100~ns have been observed \cite{Weisbuch77, Dzhioev97, Kikawa}. Long spin relaxation times in the order of microseconds have been also observed in
strong longitudinal magnetic fields $B
> 2$~T \cite{Colton, Yamamoto}.

In p-type GaAs fast electron spin relaxation on the order of several ns or shorter takes place due to electron - hole exchange interaction known as Bir-Aronov-Pikus
(BAP) mechanism \cite{BAP}. For this reason p-type structures were out of the focus until recent publications of Astakhov et al. \cite{Astakhov} in bulk and Myers
et al. \cite{Awschalom08} in quantum wells doped with magnetic Mn-acceptors. Both groups reported surprisingly long spin memory in p-type GaAs:Mn evaluated from
Hanle measurements. However the origin of this effect was interpreted differently. In Ref.~\cite{Astakhov} it was attributed to long-lived electron spin memory,
while in Ref.~\cite{Awschalom08} to the spin dynamics of Mn acceptors. This difference is surprising as the hole localization at the acceptor ($\sim$~1~nm) is
smaller than the quantum well width, so that $A^0_{Mn}$ should mainly keep its bulk properties. The problem originates from the separation of contributions of the
electron and magnetic acceptor spin orientations into the polarization of the optical $e-A^0_{Mn}$ transition, being present in any Mn-doped GaAs structure - bulk,
quantum well, wire or dot. In previous publications \cite{Astakhov, Awschalom08} the authors used indirect measurements like Hanle effect under cw excitation, where
the g-factor of the oriented particles is unknown, as well as time-resolved Kerr rotation, where the probe beam energy does not necessarily corresponds to the
energy of the transition contributing to the detected signal. Therefore both methods can not unambiguously solve the aforementioned problem.

In this letter we report on direct measurement of the spin dynamics based on polarization- and time- resolved studies of the photoluminescence (PL). This enables us
to measure g-factor, lifetime and spin relaxation time at the $e-A^0_{Mn}$ optical transition energy. From spin quantum beats with a frequency corresponding to the
electron g-factor we conclude that the long spin relaxation time in bulk GaAs:Mn is related to the electrons and not the magnetic acceptors. A rather weak
longitudinal magnetic field $B$ of about 100~mT stretches $\tau_S$ by a factor 50 from 20~ns to 1~$\mu$s, much longer than previously reported for spin relaxation
times in bulk GaAs subject to weak magnetic fields. This result is interpreted as suppression of the BAP spin relaxation mechanism for the magnetic Mn acceptor. The
data are consistent with time-resolved Faraday rotation measurements.

\begin{figure}
 \begin{minipage}{8.2cm}
  \epsfxsize=8 cm
  \centerline{\epsffile{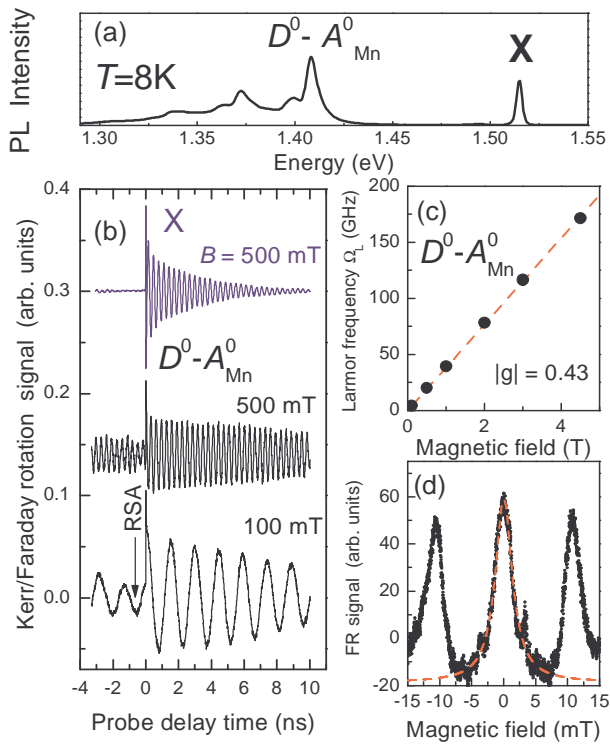}}
  \caption{\label{fig:FaradayRotation}
(a) PL spectrum of GaAs:Mn under moderate excitation with pulse power $P=1~\mu$J/cm$^2$. (b) Top: Kerr rotation at exciton resonance
$\hbar\omega_{pump}=\hbar\omega_{probe}=1.516$~eV  for degenerate pump-probe. Mid and bottom: Two-color FR with pump at 1.520~eV and probe at 1.406~eV. (c) Larmor
frequency in FR vs magnetic field. Line is linear fit to the data. (d) Resonant spin amplification in FR measured for pump-probe pulse delay $-640$~ps [see arrow in
(b)]. Dashed curve is Lorentzian fit of central peak.}
  \end{minipage}
\end{figure}

The investigated sample was grown by liquid phase epitaxy  on a (001)-oriented GaAs substrate. The thickness of the GaAs:Mn layer is $36~\mu$m and the concentration
of Mn acceptors is $8\times10^{17}$cm$^{-3}$. The acceptors are partially compensated by residual donors \cite{Astakhov}. The PL spectrum of this sample in
Fig.~\ref{fig:FaradayRotation}(a) consists of the exciton (X) and donor-acceptor ($D^0-A^0_{Mn}$) emission lines.

For Faraday (Kerr) rotation measurements  two synchronized mode-locked Ti:Sa lasers operating at independently variable wavelengths were used as sources of pump and
probe pulses. The lasers emitted pulses with 1~ps duration and 2~nm spectral width at a repetition frequency of 76~MHz. The sample was held in a split-coil cryostat
for magnetic fields $B\leq 7$~T. The laser beams were directed along the sample growth axis ($z$-axis) and the magnetic field was applied perpendicular to it,
$\mathbf{B} \perp \mathbf{z}$ (Voigt geometry). To exclude dynamic nuclear polarization, the helicity of the pump beam was modulated with a photo-elastic modulator
at 50~kHz frequency. The rotation angle of the linearly polarized transmitted (reflected) probe pulse was homodyne detected with a balanced photodiode and a lock-in
amplifier.

For time-resolved PL measurements  we used a Ti:Sa laser combined with a pulse picker, by which the pulse separation was extended to 1.3~$\mu$s. The energy of the
circularly polarized excitation photons was tuned to $\hbar\omega_{exc}=1.56$~eV, slightly above the GaAs band gap. The sample was mounted in a He-bath cryostat and
magnetic fields up to 0.2~T were applied in Voigt or Faraday geometry using an electromagnet. The PL signal was dispersed by a single monochromator with 6.28~nm/mm
linear dispersion and detected with a streak camera allowing for a time resolution down to 50~ps. PL detection with either $\sigma^+$ or $\sigma^-$ polarization
after $\sigma^+$ excitation was selected by rotating a $\lambda/4$ plate with a subsequent Glan-Thomson prism. The degree of circular polarization was determined as
$\rho_c=(I_{+}-I_{-})/(I_{+}+I_{-})$, where $I_{+}$ and $I_{-}$ are the $\sigma^+$ and $\sigma^-$ polarized PL intensities, respectively. For all experiments the
temperature was kept at $T = 8$~K.

First, we discuss results of time-resolved Faraday (Kerr) rotation. The typical signals in Fig.~\ref{fig:FaradayRotation}(b) show oscillations with a decreasing
amplitude with increasing delay between pump and probe. The signal is proportional to the time evolution of the $z$-projection of the photoinduced average spin
which can be well described by the form $\cos(\Omega_L t)\exp{(-t/T_2^*)}$. Here $T_2^*$ is the dephasing time of the spin ensemble and $\Omega_L=g\mu_B B/\hbar$ is
the Larmor frequency \cite{Crooker}, where $\mu_B$ is the Bohr magneton and $g$ is the g-factor. The value of $|g|=0.43\pm 0.01$ as determined from a linear fit to
the $\Omega_L(B)$ dependence measured for the $D^0-A^0_{Mn}$ transition (see Fig.~\ref{fig:FaradayRotation}(c)) is identical with the electron g-factor in GaAs.
This allows us to assign the measured signals to optically oriented electrons. The photogenerated holes rapidly loose their spin due to the complex valence band
structure and therefore give no contribution to the long-lived spin dynamics \cite{OptOrient}. We do not find any evidence of a Mn acceptor spin orientation, which
would give rise to oscillations with frequencies corresponding to $g_{A^0}=2.74$ and $g_{A^-}=2.02$ for the neutral and ionized Mn-acceptor, respectively
\cite{Schneider87,Sapega2001}.

For degenerate Kerr rotation,  where the energies of pump $\hbar\omega_{pump}$ and  probe $\hbar\omega_{probe}$ coincide, signal is detected only close to the
exciton (X) resonance [upper curve in Fig.~\ref{fig:FaradayRotation}(b)], decaying with the exciton lifetime $\tau_{X} = 3$~ns \cite{Elliot} and oscillating
according to $|g|=0.43$. In two-color Faraday rotation (FR) the pump laser excites free carriers at the band edge of GaAs ($\hbar\omega_{pump}=1.520$~eV), while the
probe at the $D^0-A^0_{Mn}$ energy ($\hbar\omega_{probe}=1.406$~eV) propagates through the sample without significant absorption. For weak magnetic fields $B \leq
500$~mT the signal decays on a much longer time scale compared to the exciton as one can see from the strong signal at negative delays. We measured its decay time
$T_2^*$ by the resonant spin amplification (RSA) technique \cite{Kikawa}, in which the FR signal was detected as function of magnetic field at a fixed negative
delay of $-640$~ps [Fig.~\ref{fig:FaradayRotation}(d)]. The central RSA peak is fitted by a Lorentzian,  whose half width at half maximum $B_{1/2}=1.6$~mT gives
$T_2^* =\hbar/g\mu_B B_{1/2}\approx17$~ns. The dephasing rate in the limit of low magnetic fields $1/T_2^*$ corresponds to the inverse spin lifetime $1/T_S=
1/\tau_S + 1/\tau$, where $\tau$ is the electron lifetime. The long dephasing time of $T_2^*\approx17$~ns indicates that the spin lifetime is related to long-lived
electrons, which eventually recombine with acceptor bound holes. However the FR spectral dependence on the probe pulse energy does not allow to attribute the
signal to a certain optical transition unambiguously, e.g. $e-A^0_{Mn}$, because the form of the signal remains the same for 1.4~eV$<\hbar\omega_{probe}<1.5$~eV. In
addition the spin lifetime $T_S$ depends on both $\tau$ and $\tau_S$, and for determining $\tau_S$ one needs to know $\tau$.

\begin{figure}
 \begin{minipage}{8.2cm}
  \epsfxsize=5.7 cm
  \centerline{\epsffile{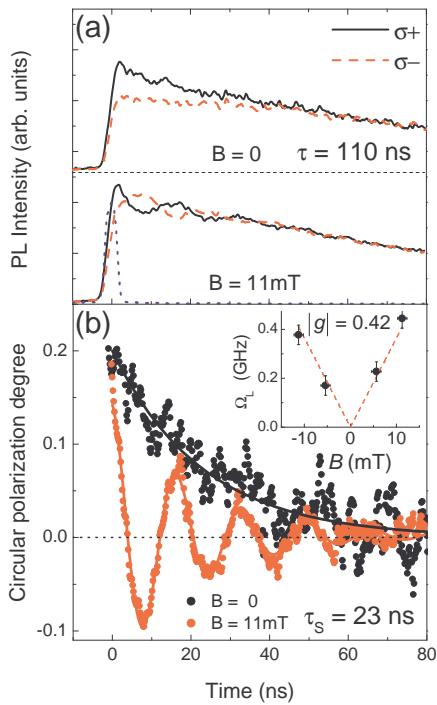}}
  \caption{\label{fig:PLtransients}
(a) PL transients of $D^0 - A^0_{Mn}$ emission line under $\sigma^+$ excitation at 1.560~eV measured for $\sigma^+$ (solid line) and $\sigma^-$ (dashed)
polarization at $B=0$ and 11~mT in Voigt geometry. The laser apparatus function is shown by the dotted curve. (b) Time evolution of circular polarization degree.
Solid lines are fits. Inset shows precession frequency $\Omega_L$ as function of magnetic field. Line is linear fit with $|g|=0.42\pm0.02$.}
  \end{minipage}
\end{figure}

Time-resolved measurements of the total PL intensity $I(t)=I_+ + I_-$ and polarization degree $\rho_c(t)$ allow us to determine both $\tau$ and $\tau_S$ exactly for
the optical $e-A^0_{Mn}$ transition energy. The intensity is related to the population decay $I(t)\propto \exp{(-t/\tau)}$, while
$\rho_c(t)=\rho_c(0)\exp{(-t/\tau_S)}$ gives direct access to the spin dynamics of oriented electrons. Moreover, in transverse magnetic field the carrier g-factor
can be determined from the oscillations of $\rho_c(t)$, similar to the pump-probe experiment \cite{Heberle94}. Transients measured at $B=0$ and 11~mT in Voigt
geometry are shown in Fig.~\ref{fig:PLtransients}. From $I(t)$ and $\rho_c(t)$ at zero magnetic field we find $\tau=110$~ns and $\tau_S=23$~ns. These values give
$T_S=19$~ns, which is in good agreement with the value obtained from FR. We do not observe a dependence of these times on excitation power up to a moderate pulse
intensity $P=1~\mu$J/cm$^2$.

In transverse magnetic field we detect pronounced oscillations of the circular polarization degree [Fig.~\ref{fig:PLtransients}(b)], which can be well fitted by
$\rho_c(t) = \rho_c(0)\cos(\Omega_Lt)\exp{(-t/\tau_S)}$. The Larmor frequency dependence on magnetic field in the inset gives an electron g-factor of
$|g|=0.42\pm0.02$. The initial electron spin is $S_z(0)=\rho_c(0)\approx0.2$, which is close to the maximum value of 0.25 \cite{OptOrient}. In agreement with the
FR, we therefore conclude from the PL data that the long-lived spin dependent signal is contributed solely by donor bound electrons.

\begin{figure}
 \begin{minipage}{8.2cm}  \epsfxsize=8 cm
  \centerline{\epsffile{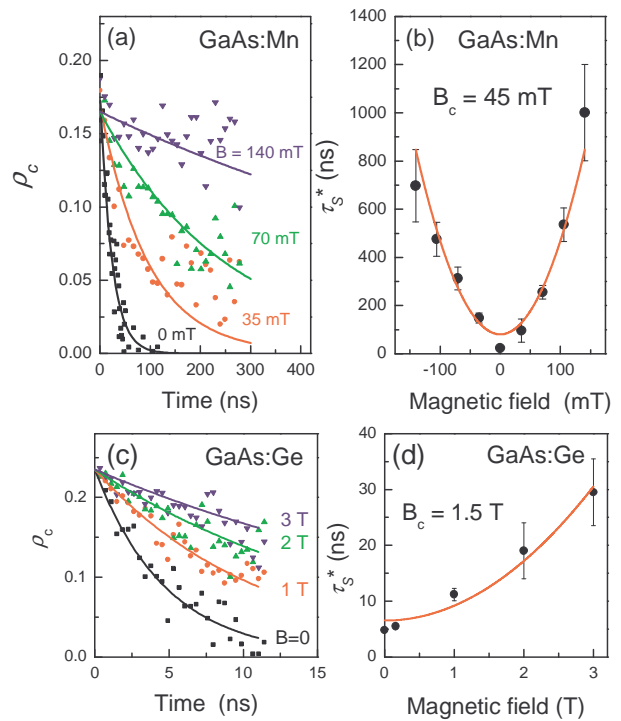}}
  \caption{\label{fig:LongitudinalField}
(a) Time-resolved circular polarization degree of $D^0-A^0_{Mn}$ emission line for different longitudinal magnetic fields at $P=0.2~\mu$J/cm$^2$. Lines are
exponential fits. (b) Spin relaxation time $\tau_S^*$ as function of magnetic field in Mn-doped sample. Line is fit with Eq.~(\ref{eq:T1-time}). (c,d) same as (a,b)
for $D^0-A^0_{Ge}$ line in GaAs:Ge sample measured at $P=1~\mu$J/cm$^2$.}
  \end{minipage}
\end{figure}

Further insight in the spin dynamics is obtained from experiments in Faraday geometry, where the electron spin relaxation is typically suppressed with increasing
magnetic field~\cite{OptOrient}. For GaAs:Mn we observe strong changes already in weak fields, see Fig.~\ref{fig:LongitudinalField}(a,b). The longitudinal electron
spin relaxation time $\tau^*_S$ increases from 20~ns at zero field up to $1~\mu$s at 140~mT, as shown by the magnetic fields dependence of $\tau^*_S$ in panel (b).

Before discussing the data let us consider the four main mechanisms responsible for electron spin relaxation. Two of them related to spin orbit interaction, the
Elliot-Yafet and Dyakonov-Perel' mechanisms, are not relevant for our case since the electrons are bound to donors at low temperatures. The hyperfine interaction of
the electron with nuclear spins can provide relaxation times in the microsecond range for donor concentrations of $10^{16}$cm$^{-3}$~\cite{Dzhioev2002}, and thus is
much longer than the typical times due to electron exchange interaction with holes bound to acceptors. This Bir-Aronov-Pikus mechanism dominates in samples with
acceptor concentrations exceeding $10^{17}$~cm$^{-3}$ \cite{BAP}.

The suppression of spin relaxation by a longitudinal magnetic field can be described in the motional averaging model by
\begin{equation}
\label{eq:T1-time}    \tau^*_S = \tau_S \left[~ 1+(B/B_c)^2
~\right].
\end{equation}
Here $B_c=\hbar/[(g_{A^0}-g)\mu_B\tau_c]$ and $\tau_c$ is the correlation time during which the random field resulting in spin relaxation can be considered as
constant \cite{OptOrient, Dyakonov1974}. The regular change of the random field due to hole spin precession in the external field is taken into account through the
neutral acceptor g-factor $g_{A^0}$ in the expression for $B_c$. The spin relaxation rate at zero magnetic field is given by:
\begin{equation}
\label{eq:T2}    \frac{1}{\tau_S} = \frac{2}{3}\omega_f^2 \tau_c.
\end{equation}
The rms value of the electron spin precession frequency $\omega_f$ in the random field characterizes the field strength. The experimental data in
Fig.~\ref{fig:LongitudinalField}(b) can be well fitted by Eq.~(\ref{eq:T1-time}), as shown by the solid line, except for the data point at $B=0$ \cite{footnote2}.
We find from this fit $\tau_S\approx70$~ns and $B_c\approx45$~mT, which is close to the continuous wave Hanle data for moderate pump intensities which give
$\tau_S=160$~ns \cite{Astakhov}. Assuming the acceptors are neutral, i.e. $g_{A^0}-g=3.16$, we find $\tau_c\approx 80$~ps.

For comparison we have measured a typical p-type GaAs:Ge sample with comparable concentration of non-magnetic Ge acceptors of $6\times10^{17}$~cm$^{-3}$ (lower
panels in Fig.~\ref{fig:LongitudinalField}). The PL transients of the donor-acceptor emission line give a decay time $\tau\approx3.3$~ns. The PL circular
polarization decays with $\tau_S=4.7$~ns at $B=0$, showing no significant change for $B<200$~mT. With increasing field it increases, and $\tau^*_S$ reaches 30~ns at
$B=3$~T~\cite{footnote1}. A fit according to Eq.~(\ref{eq:T1-time}) gives $B_c\approx1.5$~T. With $g_A=0.7$ for Ge~\cite{Sapega}, we find $\tau_c \approx 7$~ps,
which is significantly shorter than in Mn-doped GaAs. Using Eq.~(\ref{eq:T2}) we find that $\omega_f^2$ is about 100 times larger in GaAs:Ge than in GaAs:Mn.

The exchange interaction energy between the electron spin $\mathbf{S}$ and total magnetic moment of the neutral Mn acceptor $\mathbf{F}$ is $H_S=-a_F (\mathbf{S}
\cdot \mathbf{F})$, where $a_F$ depends on the exchange constant $\Delta_{A^0_{Mn}}$ and the overlap of the electron and acceptor hole wavefunctions
\cite{Astakhov}. The extension of the donor-bound electron given by the Bohr radius $a_B$ is much larger than the Mn acceptor. Therefore the precession frequency of
the electron spin in the effective magnetic field of the Mn acceptor, which is located at a distance $R$ from the donor, is $\vec{\omega}\mathbf{_f}=
\Delta_{A^0_{Mn}} \mathbf{F}\exp{(-2R/a_B)}/\hbar$. By averaging $\langle...\rangle$ over the donor-acceptor distances one obtains
\begin{equation}
\label{eq:FluctRandField}    \omega_f^2 =
\frac{\Delta_{A^0_{Mn}}^2 F(F+1)}{\hbar^2}\left\langle
\exp{\left(-\frac{4R}{a_B} \right)} \right\rangle.
\end{equation}
The Mn-acceptor ground state corresponds to the antiparallel configuration of the hole and Mn spins with angular momentum $F=1$. It follows from
Eq.~(\ref{eq:FluctRandField}) that $\omega_f^2$ depends not only on the exchange constant $\Delta_{A^0_{Mn}}$ but also on the distribution of the impurities. In
case of a non-magnetic acceptor $\omega_f^2$ is given by the same Eq.~(\ref{eq:FluctRandField}), the exchange constant $\Delta_{A^0_{Mn}}$ and momentum $F$,
however, has to be substituted by the electron-hole exchange energy $\Delta_{A^0_{Ge}}$ and the total angular momentum of the hole  $J=3/2$ \cite{Dyakonov1974}.

As mentioned, $\omega_f^2$ differs by two orders of magnitude in  Ge- and Mn-doped samples. Two reasons may be responsible for its strong decrease in GaAs:Mn. The
first is the antiferromagnetic coupling between the hole- and Mn- spins, which reduces $\Delta_{A^0_{Mn}}$ and $F$ compared with the non-magnetic Ge-acceptor
\cite{Astakhov, Dietl}. The second is related to the larger average distance $R$ between donors and acceptors in GaAs:Mn as compared to GaAs:Ge with similar doping
concentration. This is supported by a factor 30 difference in the emission decay times measured for the $D^0-A^0$ transitions.

The correlation times $\tau_c$ also differ strongly in Ge and Mn doped samples. For electrons localized on shallow donors, $\tau_c$ is determined by either the
electron spin hopping time between donors or by the spin relaxation time of the hole bound to an acceptor, whichever is the shorter one. For a donor concentration
of $10^{16}$cm$^{-3}$ the electron hoping time is on the order of 10-100~ps \cite{Dyakonov1974, Dzhioev2002}. Therefore we conclude that the short correlation time
of $\tau_c\approx7$~ps in GaAs:Ge is mainly due to hole spin relaxation. In contrast, the long $\tau_c \approx $~80~ps in GaAs:Mn suggests that the spin relaxation
of the Mn acceptor occurs on a time scale comparable or longer than the spin hopping time. The reason for this difference is the strong spin coupling of the hole
with the Mn spin.

In conclusion we have shown that the long spin relaxation time exceeding 20~ns in GaAs doped with Mn is related with electrons. The long correlation time leads to
the small external magnetic fields required to suppress the Bir-Aronov-Pikus mechanism, which controls the spin dynamics of bound electrons. An important role plays
also g-factor of the hole bound to the acceptor which is almost 4 times larger in GaAs:Mn as compared with non-magnetic GaAs. These features allows us to extend the
electron spin relaxation time up to 1~$\mu$s in weak longitudinal magnetic fields in the 100~mT range.

The authors thank G.~V.~Astakhov for useful discussions. This work was supported by the Deutsche Forschungsgemeinschaft (Grant No. 436RUS$113/958/$0-1) and  Russian
Foundation for Basic Research.


\begin{references}
\bibitem{footnote2} The three-fold reduction of $\tau_S$ from 70 to 23~ns
at zero magnetic field cannot be related to dynamical nuclear polarization, as the 20~ns time has been measured by pump-probe FR with modulated light polarization,
to suppress nuclear effects. No singularity of $\tau_S$ is observed for GaAs:Ge at zero field, indicating that the reduction is related to specifics of the magnetic
Mn acceptor. The detailed study of this effect will be performed in future.
\bibitem{footnote1} For stronger magnetic fields we analyze the average of $\rho_c(t)$ measured in opposite field directions. This allows
us to exclude effects related to carrier thermalization on the
Zeeman sublevels.
\end{references}
\end{document}